\documentclass{llncs}
\usepackage{makeidx}
\usepackage{amssymb}

\newcommand{\zah}{{\mathbb{Z}}}

\title{Constant-Depth Frege Systems with Counting Axioms Polynomially Simulate
Nullstellensatz Refutations}
\author{Russell Impagliazzo\thanks{Research Supported by NSF Award CCR-9734911,
grant $\# 93 025$ of
the joint US-Czechoslovak Science and Technology Program,
NSF Award CCR-0098197,
and USA-Israel BSF Grant 97-00188}
and  Nathan Segerlind\thanks{Partially supported by NSF grant DMS-9803515}}
\institute{Department of Computer Science \\
University of California, San Diego\\
La Jolla, CA 92093\\
\email{\{russell,nsegerli\}@cs.ucsd.edu}}

\begin{document}
\maketitle

\pagestyle{headings}

\begin{abstract}
We show that constant-depth Frege systems with counting axioms modulo $m$
polynomially simulate Nullstellensatz refutations modulo $m$.  
Central to this is a new definition of reducibility from formulas to systems of
polynomials with the property that, for most previously studied translations of
formulas to systems of polynomials,  a formula reduces to its translation.
When combined with a previous result of the authors,  this establishes the
first size separation between Nullstellensatz and polynomial calculus 
refutations.  We also obtain new, small refutations for certain
CNFs by constant-depth Frege systems with counting axioms. 
\end{abstract}

\section{Introduction}
This paper studies proof sizes in propositional systems that utilize modular 
counting in limited ways.   The complexity of propositional proofs has 
received much attention in recent years because of its connections to 
computational and circuit complexity 
\cite{cookReckhow,krajicek95,pudlakHandbook,BP98:propcp}.  In particular,
$NP$ equals $coNP$ if and only
if there exists a propositional proof system that proves every tautology in 
size polynomial in the size of the tautology  \cite{cookReckhow}.
But before we can prove lower bounds for all proof systems,  it seems
necessary that we be able to prove lower bounds for specific proof systems.
There was much initial success showing lower bounds for constant-depth proof
systems \cite{ajtai88,KraPudWoo95,CC::PitassiBI1993}.  While these proof 
systems can simulate many powerful theorem proving techniques,  such as 
resolution,  they cannot perform reasoning that involves modular counting.
For this reason,  there has been much interest in recent years regarding
proof systems that incorporate modular counting in different ways.  Three such
systems are: constant-depth Frege systems augmented with counting axioms 
\cite{Ajtai90,Ajtai1994,APAL::BeameP1996,FOCS::BeameIKPP1994,BussEtAl97,Riis1997,beameRiis97,is01sepbfcpc} (counting axioms state that a set of size $N$ cannot
be partitioned into sets of size $m$ when $N$ is indivisible by $m$),
the Nullstellensatz system 
\cite{FOCS::BeameIKPP1994,BussEtAl97,beameRiis97,bussPitNullBounds,homoPoly99},
which captures static polynomial reasoning,  and the polynomial calculus
\cite{CleggEI1996,Razborov98,ImpPudSga99,bensassonImpagliazzo99,bgip99},
which captures iterative polynomial reasoning.

We show that constant-depth Frege systems with counting  axioms modulo $m$ 
polynomially simulate Nullstellensatz refutations modulo $m$.    This allows us
to transform Nullstellensatz refutations into constant-depth Frege with 
counting axioms proofs with a small increase in size,
and to infer size lower bounds for Nullstellensatz refutations from
size lower bounds for  constant-depth Frege with counting axioms proofs.
In particular, this method
establishes the first superpolynomial size separation between
Nullstellensatz and polynomial calculus refutations.

Our simulation also shows that previously used proof techniques were not
only sufficient but necessary.   
Papers such as \cite{FOCS::BeameIKPP1994,BussEtAl97,beameRiis97,is01sepbfcpc}
prove size lower bounds for constant-depth Frege systems with counting axioms
by converting
small proofs into low degree  Nullstellensatz refutations. The
existence of such low degree  Nullstellensatz refutations is then
disproved by algebraic and combinatorial means.
Low degree Nullstellensatz refutations are small (because there are few
monomials), so our simulation shows that if there were such a low degree
Nullstellensatz refutations,  there would be a small constant-depth Frege with
counting axioms proof.  Therefore,  Nullstellensatz degree lower bounds
are necessary for size lower bounds for constant-depth
Frege systems with counting axioms.

It is not immediately clear how to compare constant-depth Frege systems with 
Nullstellensatz refutations because Frege systems prove propositional formulas
in connectives such as $\bigwedge$, $\bigvee$ and $\neg$,  and the
Nullstellensatz system shows that systems of polynomials have no
common roots.
We propose a new definition of reducibility from 
propositional formulas to systems of polynomials:  a formula $F$ reduces to a 
system of polynomials over $\zah_m$ if we can use $F$ to define an 
$m$-partition (a partition in which every class consists of exactly $m$
elements) on the satisfied monomials of the polynomials.  The simulation shows
that if a formula has a small reduction to a set of polynomials with a small
Nullstellensatz refutation,  then the formula has a small refutation in 
constant-depth Frege with counting axioms.  
This notion of reduction seems
natural in that for previously studied translations of formulas into systems
of  polynomials,  a formula reduces to its translation.

\subsection{Outline of the Paper}

In section \ref{notationSec},  we give some definitions that we will we use in
the rest of the paper.

The simulation of Nullstellensatz refutations modulo $m$ by constant-depth 
Frege systems with counting axioms modulo $m$ works by defining two different 
$m$-partitions on the satisfied monomials in the expansion of the 
Nullstellensatz refutation.  One covers the satisfied monomials perfectly,
and the  other leaves out exactly one satisfied monomial.  In section
\ref{contraryPart},  we show that  Frege systems  with counting axioms 
can prove in constant depth and polynomial size that such a partition can not
exist.

Section \ref{simulation} formalizes our definition of reducibility from
propositional formulas to systems of polynomials and proves the main simulation
theorem.

In section \ref{uniformity} we show that, for 
several methods of translating propositional formulas into systems of
polynomials,  a formula efficiently reduces to its translation.

We explore some applications of the simulation in section \ref{linEq}.
First,  we obtain small constant-depth Frege with counting axioms refutations
for unsolvable systems of linear equations in which each equation contains
a small number of variables.  This class of tautologies includes the Tseitin
tautologies and the ``$\tau$ formulas'' for Nisan-Wigderson pseudorandom 
generators built from the parity function 
\cite{Alekhnovich:2000:PGP,krajicek:2001a}.  The Tseitin tautologies
on a constant degree expander can be expressed as an unsatisfiable set of
constant-width clauses,  and are known to 
require exponential size to refute in constant-depth Frege systems
\cite{BenSasson02a}.  Therefore,  as a corollary, 
we obtain an exponential separation of constant-depth Frege systems with 
counting axioms and constant-depth Frege systems with respect to constant-width
CNFs.

\section{Definitions, Notation and Conventions}{\label{notationSec}}

In this paper,  we perform many manipulations on partitions of sets into
pieces of a fixed size. We make use of  the following definitions:

\begin{definition}
Let $S$ be a set. The set {\em{${[S]}^m$}} is the collection of $m$ 
element subsets of $S$;
${[S]}^m  = \{ e \mid e \subseteq S, \ |e| = m \}$.
For $e,f \in {[S]}^m$,  we say that {\em{$e$ conflicts with 
$f$,  $e \perp f$,}}
 if $e \neq f$ and $e \cap f \neq \emptyset$.

\end{definition}

When $N$ is a positive integer,  we write  $[N]$ for the set
of integers $\{ i \mid 1 \leq i \leq N \}$.   The collection of $m$ element
subsets of $[N]$ are denoted by ${[N]}^m$,
{\em{not}} by ${[[N]]}^m$.

Throughout this paper, we use the word polynomial to mean ``multivariate
polynomial.''

\begin{definition}
A {\em{monomial}} is a product of variables.  A {\em{term}} is scalar multiple
of a monomial.
\end{definition}

\begin{definition}
For a monomial $t = \prod_{i \in I}x_i^{\alpha_i}$, its 
multilinearization,  {\em{${\bar{t}}$}}, is defined
as $ {\bar{t}} = \prod_{i \in I}x_i$.
Let $f = \sum_{t} c_t t$ be a polynomial. The 
multilinearization of $f$, ${\bar{f}}$, 
is defined as  ${\bar{f}} = \sum_t c_t {\bar{t}}$.
We say that a polynomial $f$ is multilinear if $f = {\bar{f}}$.
\end{definition}

\begin{definition}
Let $n >0$ be given, and let $x_1, \ldots, x_n$ be variables.  Let 
$I \subseteq [n]$ be given.  The monomial {\em{$x_I$}} is defined to be 
$\prod_{i \in I} x_i$.
\end{definition}

Notice that a multilinear polynomial $f$ in the variables $x_1, \ldots, x_n$ 
can be written as $\sum_{I \subseteq [n]} a_I x_I$.

\subsection{Proof Systems}

Propositional proof systems are usually viewed as deriving tautologies by 
applying inference rules to a set of axioms.  However,  it can be useful to
take the dual view that such proof systems establish that  a set of hypotheses
is unsatisfiable by deriving FALSE from the hypotheses and axioms.
Such systems are called refutation systems.
The Nullstellensatz and polynomial calculus systems demonstrate that sets of
polynomials have no common solution,  and are inherently refutation systems.
Frege systems are traditionally viewed as deriving tautologies,
but for ease of comparison, we treat them as refutation systems.

Furthermore,  we will be discussing propositional formulas and polynomials in
the same set of variables.  This is justified by identifying the logical 
constant FALSE with the field element $0$ and the logical constant TRUE with
the field element $1$.

\subsubsection{Constant-Depth Frege Systems} 

A Frege system is a sound,  implicationally complete propositional
proof system over a finite  set of connectives with a finite number of axiom
schema and inference rules. By the methods of Cook and Reckhow
 \cite{cookReckhow},  any
two Frege systems simulate one another up to a polynomial factor in  size and
a linear factor in depth.  For concreteness,   the reader can keep in mind
the  following Frege system whose connectives are NOT gates, $\neg$,
 and unbounded fan-in OR gates, $\bigvee$,  and whose inference rules are:  
(1) Axioms ${\overline{A \vee \neg A}}$,
(2) {Weakening} $\frac{A}{A \vee B}$
(3) {Cut} $\frac{ A \vee B \ \ (\neg A) \vee C}{B \vee C}$
(4) {Merging}$\frac{\bigvee X \vee \bigvee Y }{\bigvee \left(X \cup Y\right)}$
(5) {Unmerging} 
$\frac{\bigvee \left( X \cup Y \right)}{ \bigvee X \vee \bigvee Y}$.

Let ${\cal{H}}$ be a set of formulas. 
A derivation from ${\cal{H}}$ is a sequence of formulas
$f_1, \ldots, f_m$ so that for each $i \in [m]$, either $f_i$ is a
substitution instance of an axiom,  $f_i$ is an element of ${\cal{H}}$,  or 
there exist $j,k<i$ so that $f_i$ follows from $f_j$ and $f_k$ by the 
application of an inference rule to $f_j$ and $f_k$.

For a given formula $F$,  a proof of $F$ is a derivation from
the empty set of hypotheses whose final formula is $F$.

For fixed set of hypotheses ${\cal{H}}$,  a refutation of 
${\cal{H}}$ is a derivation from ${\cal{H}}$ whose final formula 
is FALSE.

The size of a derivation is the total number of symbols appearing in 
it.

We say that a family of tautologies $\tau_n$,  each of size $s(n)$,
has polynomial size constant-depth Frege proofs (refutations) 
if there are  constants $c$ and $d$ so that for all $n$, there is a proof
(refutation) of $\tau_n$ so that each formula in the proof has depth at most
$d$, and the proof (refutation) has size $O \left( s^c (n) \right)$.

\subsubsection{Counting Axioms Modulo $m$}

Constant-depth Frege with counting axioms modulo $m$
is the extension of constant-depth Frege
systems that has  axioms that state
for integers $m,N$,  $m \geq 2$ and  $N \not\equiv_m 0$, it is impossible to
partition a set of $N$ elements into pieces of size $m$.

\begin{definition}
Let $m >1$ and $N \not\equiv_m 0$ be given. Let $V$ be a  set of $N$ elements.
For each $e \in {[V]}^m$, let there be a variable $x_e$.

\[  {\mbox{Count}}^V_m =  \bigvee_{v \in V} \left( \bigwedge_{e \in {[V]}^m 
\atop e \ni v} \neg x_e \right) \ \ \vee \bigvee_{ e,f \in {[V]}^m \atop e
 \perp f} \left( x_e \wedge x_f \right) \]

\end{definition}

Frege with counting modulo $m$ derivations are Frege derivations that
allow the use of substitution instances of ${\mbox{Count}}^{[N]}_m$ 
(with $N \not\equiv_m 0$) as axioms.

\subsubsection{Nullstellensatz Refutations}

One way to prove that a system of polynomials $f_1, \ldots, f_k$ has no
common roots is to give a list of polynomials $p_1, \ldots, p_k$ so that 
$\sum_{i=1}^k p_i f_i = 1$.  
Because we are interested in translations of propositional formulas, we add
the polynomials $x^2 -x$ as hypotheses to guarantee all roots are 
zero-one roots.

\begin{definition}
For a system of polynomials $f_1, \ldots, f_k$ in variables $x_1, \ldots, x_n$
over a field $F$, a {\em{Nullstellensatz refutation of $f_1, \ldots, f_k$}}
 is a list of
polynomials $p_1, \ldots, p_k$, $r_1, \ldots, r_n$ satisfying the following
equation:
\[ \sum_{i=1}^k p_i f_i + \sum_{j=1}^n r_j  \left(x_j^2 - x_j \right) = 1\]

For a polynomial $q$,  a {\em{Nullstellensatz derivation of $q$ from
$f_1, \ldots, f_k$}} is  a list of
polynomials $p_1, \ldots, p_k$, $r_1, \ldots, r_n$ satisfying the following
equation:
\[ \sum_{i=1}^k p_i f_i + \sum_{j=1}^n r_j  \left(x_j^2 - x_j \right) = q\]

\end{definition}

The degree of the refutation (derivation) is the maximum degree
of the polynomials $p_i f_i$, $r_j \left( x_j^2 - x_j \right)$.

We define the size of a Nullstellensatz refutation (derivation) to
be the number of monomials appearing in $p_1, \ldots, p_k$ and 
$f_1, \ldots, f_k$.

Hilbert's weak Nullstellensatz guarantees that over a field, all unsatisfiable
systems of polynomials have Nullstellensatz refutations \cite{pitassiAPPS97}. 
We can define Nullstellensatz refutations over any ring, but such systems are
no longer complete.  In this paper,  we work with Nullstellensatz
refutations of polynomials over  $\zah_m$,  and for
the sake of generality,  we make no assumptions on $m$ unless otherwise
stated.

\subsubsection{Polynomial Calculus}

\begin{definition}
Let $f_1, \ldots,  f_k$ be polynomials over a field $F$.
A {\em{polynomial calculus refutation of $f_1, \ldots, f_k$ over $F$}} is a
sequence of polynomials
$g_1, \ldots, g_m$ so that, $g_m = 1$, and  for each $i \in [m]$, either 
$g_i$ is  $f_l$ for some $l \in [k]$,  $g_i$ is ${x_l}^2 - x_l$ for some 
$l \in [n]$,  $g_i$ is  $ag_j + bg_l$ for some
$j,l < i, a,b \in F$,  or $g_i$ is $x_l g_j$ for some $j<i$, $l \in [n]$.
\end{definition}

The size of a polynomial calculus refutation is the total number of 
monomials appearing in the polynomials of the refutation.
The degree of a polynomial calculus refutation
is the maximum degree of a polynomial that appears in the refutation.

\section{Contradictory Partitions of Satisfied Variables}{\label{contraryPart}}

To simulate Nullstellensatz refutations in constant-depth Frege systems
 with counting
axioms, we construct two partitions on the satisfied monomials of the 
refutation: one which covers the satisfied monomials exactly,  and another 
which covers the satisfied monomials with $m-1$ new points.  This is 
impossible,  and in this section,  we show that constant-depth Frege systems
with counting axioms can prove that this is impossible with polynomial
size proofs.

\begin{definition}
Let positive integers $n$ and $k$ be given.
Let $u_1, \ldots,  u_n$ be a set of Boolean variables.
For each $e \in {\left[ n \right]}^m$, let $y_e$ be a variable, and
for each $e \in {\left[  n +  k \right]}^m$,  let $z_e$ be a variable.
{\em{${\mbox{CP}}^{n,k}_m( {\vec{u}}, {\vec{y}},{\vec{z}})$}}
is the  negation of the conjunction of the following formulas:

\begin{tabbing}
123\=123\=123\=123\=123\=123\=123\=123\=123\=123\=\kill
  ``every variable covered by the first partition is satisfied'' \\
\> for each $e \in [n]^m$,  $y_e \rightarrow \bigwedge_{i \in e} u_i$ \\
   ``every satisfied variable is covered by the first partition'' \\
\> for each $i \in [n]$,  $u_i \rightarrow \bigvee_{e \ni i} y_e $ \\
   ``no two overlapping edges are used by the first partition'' \\
\> for each $e,f \in {[n]}^m$ with $e \perp f$,  $\neg y_e \vee \neg y_f $ \\
  ``every variable covered by the second partition is satisfied'' \\
\> for each $e \in {[n+k]}^m$,  
  $z_e \rightarrow \bigwedge_{i \in e \atop i \leq n} u_i$ \\
  ``every satisfied variable is covered by the second partition'' \\
\> for each $i \in [n]$,  $u_i \rightarrow \bigvee_{e \ni i} z_e$ \\
  ``every extra point is covered by the second partition'' \\
\> for each $i$, $n+1 \leq i \leq n+k$, $\bigvee_{e \ni i} z_e $ \\
   ``no two overlapping edges are used by the second partition'' \\
\> for each $e,f \in {[n+k]}^m$ with $e \perp f$,  $ \neg z_e \vee \neg z_f$
\end{tabbing}

\end{definition}

\begin{lemma}{\label{contPart}} Fix $m$ and $k$ so that
$m$ is not divisible by $k$.  For all $n$,  the tautology
${\mbox{CP}}^{n,k}_m$ has a constant depth, size $O(n^m)$ proof in
constant-depth Frege with counting modulo $m$ axioms.
\end{lemma}

\begin{proof}
Fix $m$, $n$ and $k$.  The proof of ${\mbox{CP}}^{n,k}_m$ is by contradiction.
We  define a set $U$ of size $mn + k$ and formulas $\phi_e$ for each
$e \in {[U]}^m$ so that we can derive
$\left(\neg {\mbox{Count}}^{U}_m\right) [x_e \leftarrow \phi_e]$ in
size $O(n^m)$ from the hypothesis $\neg {\mbox{CP}}^{n,k}_m$.

Let $U$ be the set consisting of the following points:
$p_{r,i}$,  $ r \in [m], \ i \in [n]$ (the $r$'th copy of the row of variables)
and $p_{m,i}$, $n + 1 \leq i \leq k$ (the extra points.) 

\begin{tabbing}
123\=123\=123\=123\=123\=123\=123\=123\=123\=123\=123\=\kill
 ``when $u_i$ is unset,  we group together its copies'' \\
\> for each $i \in [n]$,   $\phi_{\{p_{1,i}, \ldots, p_{m,i}\}} =  \neg u_i$ \\
 ``in the first $m-1$ rows, use the partition given by the $y_e$'s'' \\
\> for each $r \in [m-1]$, each $i_1, \ldots, i_m \in [n]$,  $\phi_{ \{p_{r,i_1}, \ldots, p_{r,i_m} \}}=  y_{\{i_1, \ldots, i_m \}}$ \\
  ``in the last row,  use the the partition given by the $z_e$'s'' \\
\> for each $i_1, \ldots, i_m \in [n+k]$,
  $\phi_{\{p_{m,i_1}, \ldots, p_{m,i_m} \}} = z_{\{i_1, \ldots, i_m \}}$ \\
other edges are not used \\
\> for all other $e \in [U]^m$,  $\phi_e = 0$ 
\end{tabbing}

Now we sketch the derivation of  $\left(\neg {\mbox{Count}}^{U}_m\right) [x_e \leftarrow \phi_e]$ from
 $\neg {\mbox{CP}}^{n,k}_m$.  It is easily verified that the derivation has constant depth and
size $O({(mn+k)}^{m}) = O(n^m)$.

``Every point of $U$ is covered by the partition.''

Let $p_{r,i} \in U$ with $i \in [n]$,  $r \in [m-1]$ be given.
From $\neg {\mbox{CP}}^{n,k}_m$  derive
$u_i \rightarrow \bigvee_{f \in {[n]}^m \atop f \ni i} y_f$.  
Because $\bigvee_{f \in {[n]}^m \atop f \ni i} y_f$ is a sub-disjunction of
$\bigvee_{e \in {[U]}^m \atop e \ni p_{r,i}} \phi_e$,
we may derive $u_i \rightarrow \bigvee_{e \in {[U]}^m \atop e \ni p_{r,i}} \phi_e$ with a weakening inference. 
Because 
$\phi_{\{p_{1,i}, \ldots, p_{m,i}\}} = \neg u_i$,  we may derive 
$\neg u_i \rightarrow \bigvee_{e \in {[U]}^m \atop e \ni p_{r,i}} \phi_e$.   Combining these two formulas
yields $\bigvee_{e \in {[U]}^m \atop e \ni p_{r,i}} \phi_e$.
The case for $p_{m,i}$, $i \in [n]$  is similar.

For a point  $p_{m,i}$, $n+1 \leq i \leq n+k$,  from $\neg {\mbox{CP}}^{n,k}_m$ derive
$\bigvee_{f \in {[n + k]}^m \atop f \ni i} z_f$ .   A weakening inference applied to this
derives $\bigvee_{e \ni p_{m,i}} \phi_e$.

``No overlapping edges are used.''

Let $e_1, e_2 \in {[U]}^m$ be given so that $e_1 \perp e_2$, and neither $\phi_{e_1}$ nor $\phi_{e_2}$
is identically $0$.  

If $\phi_{e_1} = \neg u_i$ and $\phi_{e_2} = y_f$,  then  $e_1$ is $\{ p_{r,i} \mid r \in [m] \}$ and
$e_2$ is $\{ p_{r,j} \mid j \in f \}$ for some $r \in [m]$ and $f \in {[n]}^m$ so that $i \in f$.
From $\neg {\mbox{CP}}^{n,k}_m$ derive $y_f \rightarrow u_i$.
From this,  derive $\neg \neg u_i \vee \neg y_f  = \neg \phi_{e_1} \vee \neg \phi_{e_2}$.  

If $\phi_{e_1} = y_{f_1}$ and $\phi_{e_2} = y_{f_2}$,  then $e_1$ is $\{p_{r_1,i} \mid i \in f_1\}$
and $e_2$ is $\{p_{r_2,i} \mid i \in f_2\}$ with $r_1 = r_2$ and $f_1 \perp f_2$.
From  $\neg {\mbox{CP}}^{n,k}_m$ derive $\neg y_{f_1} \vee \neg y_{f_2} = \neg \phi_{e_1} \vee \neg \phi_{e_2}$.

The only other cases are when $\phi_{e_1} = \neg u_i$ and $\phi_{e_2} = z_f$
or $\phi_{e_1} = z_{f_1}$ and $\phi_{e_2} = z_{f_2}$,  and these are handled similarly. 

\end{proof}

\section{The Simulation}{\label{simulation}}

Because we work over $\zah_m$,  a polynomial  vanishes on a given assignment if and only if there is an
$m$-partition on its satisfied monomials  (recall that we treat a monomial with coefficient
$a$ as having $a$ distinct copies.)
The definability of this partition is the connection between refuting a propositional
formula and refuting a system of polynomials.   

\subsection{Reducing Formulas to Systems of Equations} \label{reductionSS}

The method we use to reduce a formula to a system of polynomials is 
to define a partition on the satisfied monomials of the polynomials with
small, constant-depth formulas and prove that these formulas define a partition
using the formula as a hypothesis.

Because of the central role played by the sets of monomials appearing in
each polynomial,  we take a moment to define this notion precisely.
First of all,  because we are concerned only 
with $0/1$ assignments,  a polynomial vanishes if and only if its 
multilinearization vanishes.  For this reason,  we restrict our attention to
multilinear polynomials.
We treat a term $a x_I$ as $a$ distinct
copies of the monomial $x_I$.  For this reason,  when we talk about the 
``set of monomials'' of a polynomial,  we do not mean the set of monomials
that appear in the polynomial, but a set which includes $a$ copies of each
monomial with coefficient $a$.
We will generally identify $a x_I$ with $a$ objects $m_{1,I}, \ldots, m_{a,I}$.
Think of $m_{c,I}$ as the $c$'th copy of the monomial $x_I$.
There should be little confusion of the dual use of the symbol 
``$m$'' because when the symbol appears without
a subscript it denotes the modulus,  and when it appear with a subscript it
denotes a monomial.

\begin{definition}{\label{reduceDefn}}
Let  $f = \sum_{I \subseteq [n]} a_I x_I$ be a multilinear polynomial over
$\zah_m$. The {\em{set of monomials of $f$}} is the following set:

\[ M_f = \{ m_{c,I}  \mid I \subseteq [n],  \ c \in [a_I]  \} \]
\end{definition}

\begin{definition}
Let $x_1, \ldots, x_n $  be Boolean variables.
Let $f$ be a multilinear polynomial in the variables $x_1, \ldots, x_n$.
For each $E \in {\left[ M_{f} \right]}^m$,  let  {\em{$\theta_E$}}
be a formula in
${\vec{x}}$.
We say that the $\theta$'s form an $m$-partition the
satisfied monomials of $f$ if the following formula holds:

\[  \bigwedge_{E \in {[M_{f}]}^m } \left(  \theta_E \rightarrow \bigwedge_{ m_{c,I} \in E  } \bigwedge_{k \in I} x_k \right)  \ \wedge  \ 
 \left( \bigwedge_{E,F \in {[M_{f}]}^m \atop  E \perp F} \neg \theta_E \vee \neg \theta_F \right) \]
\[ \wedge \  \bigwedge_{ m_{c,I} \in M_{f} }\left( \left( \bigwedge_{k \in I}  x_k \right) \rightarrow   \bigvee_{E \in {[M_{{f}}]}^m \atop E \ni  m_{c,I}  } \theta_E  \right)  \]
\end{definition}

\begin{definition}
Let $x_1, \ldots, x_n$ be  Boolean variables.
Let $\Gamma({\vec{x}})$ be a propositional formula.
Let $F = \{ f_1, \ldots, f_k \}$ be a system of polynomials over $\zah_m$
with a Nullstellensatz refutation $p_1, \ldots, p_k$, $r_1, \ldots, r_n$.
If,  for each $i \in [k]$,
there are formulas $\beta^i_E({\vec{x}})$, 
$E \in { \left[ M_{\bar{f_i}} \right]}^m$,   so
that there is a size $T$, depth $d$ Frege derivation from 
$\Gamma({\vec{x}})$ that,  for each $i$, the
$\beta^i$'s form an $m$-partition on the satisfied monomials of ${\bar{f_i}}$,
then we say that {\em{$\Gamma$ reduces to  $F$ in depth
$d$ and size $T$}}.

\end{definition}

\subsection{The Simulation}\label{simulationSS}
\begin{theorem}{\label{theSimulation}}
Let $m>1$ be an integer.
Let $x_1, \ldots, x_n$ be  Boolean variables.
Let $\Gamma({\vec{x}})$ be a propositional formula,  and
let $F$ be a system of polynomials over $\zah_m$
so that $\Gamma$ reduces to $F$ in depth $d$ and size $T$.
If there is a Nullstellensatz refutation of $F$ with 
size $S$,  then there is a depth $O(d)$ Frege with counting axioms modulo $m$ 
refutation  of $\Gamma({\vec{x}})$ with size  $O(S^{2m}T)$.
\end{theorem}

\begin{proof}
Let $p_1, \ldots, p_k$, $r_1, \ldots, r_n$ be a size $S$ Nullstellensatz
refutation of $F$.  Let $\beta^i_E({\vec{x}})$,  for $i \in [k]$, 
$E \in {[M_{{\bar{f_i}}}]}^m$,  be  formulas so that from $\Gamma$ there is a 
size $T$, depth $d$ proof  that for each $i$ the $\beta^i_E({\vec{x}})$'s form
an $m$-partition on the satisfied monomials of ${\bar{f_i}}$.

We obtain contradictory partitions of the the monomials that appear in
the expansion of $\sum_{i=1}^k {\bar{p_i}} {\bar{f_i}}$ 
in which polynomials are multiplied and multilinearized, but no terms are
collected.  In other words, the set is the collection, over $i \in [k]$,
of all pairs of monomials from $\bar{p_i}$ and $\bar{f_i}$.

\[ V = \bigcup_{i=1}^k \{ \left( m_{c,I},m_{d,J}, i \right) \mid m_{c,I} \in M_{{\bar{p_i}}}, \ m_{d,J} \in M_{{\bar{f_i}}} \} \]
Notice that $|V| = O(S^2)$.

For each $v \in V$,  $v = \left( m_{c,I},m_{d,J}, i \right)$,  let
$\gamma_v = \bigwedge_{k \in I \cup J} x_k$.
Think of these as the monomials.
We will give formulas
$\theta_E$, that define a partition on the satisfied monomials with $m-1$ many extra points,
and $\eta_E$,  that define a partition on the satisfied monomials with no extra points.  We will give a
$O({|V|}^m + T) = O( S^{2m} + T  )$ derivation from $\Gamma$ of the following:

\[\neg {\mbox{CP}}^{|V|,m-1}_m  \left[u_v \leftarrow \gamma_v, \ y_E \leftarrow \theta_E, \ z_E \leftarrow \eta_E   \right] \]

On the other hand,  by lemma \ref{contPart}, ${\mbox{CP}}^{|V|,m-1}_m$
has constant depth Frege proofs of size
$O({|V|}^m)$,  so
${\mbox{CP}}^{|V|,m-1}_m \left[u_v \leftarrow \gamma_v, \ 
y_E \leftarrow \theta_E, \ z_E \leftarrow \eta_E   \right]$
has a constant depth Frege proof of size $O({|V|}^m T)$.
Therefore,  $\Gamma$ has  a depth $O(d)$ Frege refutation of size $O(S^{2m}T)$.

{\bf{The Partition with  $m-1$ Extra Points}}

Notice that we have the following equation:
\[{\overline{\sum_{i=1}^k {\bar{p_i}} {\bar{f_i}}}} =
{\overline{\sum_{i=1}^k p_i f_i + \sum_{j=1}^n r_j(x^2_j - x_j)}} = 1\]
So when we collect terms after expanding $\sum_{i=1}^k {\bar{p_i}} {\bar{f_i}}$
and multilinearizing,  the coefficient of every nonconstant term is  $0$ 
modulo $m$, and the constant term is $1$ modulo $m$. 

For each $S \subseteq \left[ n \right]$,  let
$V_S = \{ \left( m_{c,I}, m_{d,J}, i)  \right) \in V \mid  I \cup J = S \}$.
Think of these as the occurrences of $x_S$ in the multilinearized expansion.

For each $S \subseteq [n]$, $S \neq \emptyset$,
there is  an $m$-partition  on 
$V_S$,   call it ${\cal{P}}_S$. Likewise,  there is an $m$-partition on
$V_\emptyset \cup [m-1]$,  call it  ${\cal{P}}_\emptyset$.

Define the formulas $\theta_E$ as follows:  for each
$E \in {\left( [V] \cup [ m-1] \right)}^m$,  if $E \in {\cal{P}}_S$
for some $S \subseteq [n]$ then $\theta_E = \bigwedge_{k \in S} x_k$, 
otherwise $\theta_E = 0$.

Constant-depth Frege can prove that this is a $m$-partition of  the satisfied monomials of $\sum_{i=1}^k {\overline{{\bar{p_i}} {\bar{f_i}}}}$ with $m-1$ extra points.  The proof
has size $O ({|V|}^m)$ and depth $O(1)$. It is trivial from the definition of $\theta_E$ that the edges cover
only satisfied monomials.  That every satisfied monomial $\bigwedge_{k \in S} x_k$ is covered 
is also trivial: the edge from
${\cal{P}}_S$ is used if and only if  the term $x_S$ is satisfied.  Finally,  it easily shown
that the formulas for two
overlapping edges are never both satisfied:  only edges from
${\cal{P}}_S$ are used (regardless of the values of the $x$'s),  so for any pair of overlapping
edges, $E \perp F$,  one of the two formulas $\theta_E$ or $\theta_F$ is identically $0$.

{\bf{The Partition with No Extra Points}}

The idea is that  an $m$-partition
on the satisfied monomials on ${\bar{f_i}}$  can be used to build an $m$-partition on the satisfied monomials of $t {\bar{f_i}}$,
for any monomial $t$.


For each $E \in {\left[ V \right]}^m$, define $\eta_E$ as follows:  if 
$E = \{ \left( m_{c,I}, m_{d_l, J_l},i \right) \mid  l \in [m] \}$ for some
$i \in [k]$, $ m_{c,I} \in M_{f_i}$, then $\eta_E  = \bigwedge_{k \in I} x_k \wedge \beta_{\{m_{d_l, J_l} \mid l \in [m] \}}$,
otherwise,  $\eta_E = 0$.

There is a size $O(S + |V|^m )$, depth $O(d)$ Frege derivation from $\Gamma$ that
the $\eta_E$'s form an $m$-partition on the satisfied monomials of $\sum_{i=1}^k {\overline{{\bar{p_i}}{\bar{f_i}}}}$.
We briefly sketch how to construct the proof.  Begin by deriving  from $\Gamma$,  for each $i$,
that the $\beta_E^i$'s  form an $m$-partition on the satisfied monomials of ${\bar{f_i}}$.

``Every satisfied monomial is covered.''
Let $\left( m_{c,I}, m_{d,J}, i \right) \in V$ be given.
If $\bigwedge_{k \in I \cup J} x_k $ holds, then  so do
$\bigwedge_{k \in I} x_k$ and $ \bigwedge_{k \in J} x_k $.  
Because the $\beta^i$'s form  an $m$-partition on the satisfied monomials of ${\bar{f_i}}$,  
we may derive $\bigvee_{F \in {\left[ M_{f_i}\right]}^m}  \beta^i_F$.  From this derive
$\bigvee_{F \in {\left[ M_{f_i}\right]}^m} \bigwedge_{k \in I} x_k \wedge \beta^i_{F}$.
A weakening inference applied to this yields
$\bigvee_{E \in {\left[ V \right]}^m} \eta_E$.

``Every monomial covered is satisfied.''
Let $v = \left( m_{c,I}, m_{d,J}, i \right) \in V$ be given so that
$v \in E$ and $\eta_E $ holds.  For this to happen,  $E = \{ \left( m_{c,I}, m_{d_l,J_l},  i \right) \mid  l \in [m] \}$.
By definition,,  $\eta_E = \bigwedge_{k \in I} x_k \wedge \beta^i_{\{ m_{d_l, J_l} \mid l \in [m]  \} }$,
and therefore
$\bigwedge_{k \in I} x_k$ holds. Because the
$\beta^i$'s form an $m$-partition on the satisfied  monomials of ${\bar{f_i}}$,
we have that $\bigwedge_{k \in J} x_k$
holds. Therefore $\bigwedge_{k \in I \cup J} x_k$ holds.

``No two conflicting edges $E$ and $F$ 
can have $\eta_E$ and $\eta_F$ simultaneously
satisfied.''
 If $E \perp F$,  and
neither $\theta_E$ nor $\theta_F$ is identically $0$,  then they share the same
${\bar{p_i}}$ component.  That is,  there exists $i$,  $m_{c,I} \in M_{{\bar{p_i}}}$ so that 
$E = \{ \left( m_{c,I}, m_{d_l,J_l},  i \right) \mid   l \in [m] \}$,
and $F = \{ \left( m_{c,I}, m_{{d'}_l,{J'}_l},  i \right) \mid  l \in [m] \}$.
Because $E \perp F$,  we have $ \{ m_{d_l,J_l} \mid l \in [m] \}  \perp  \{ m_{{d'}_l,{J'}_l} \mid l \in [m] \} $.
Because the $\beta^i$'s form an $m$-partition on the
satisfied monomials of ${\bar{f_i}}$,  we can derive $\neg \beta^i_{ \{ m_{d_l,J_l} \mid l \in [m] \}} \vee \neg \beta^i_{ \{ m_{{d'}_l,{J'}_l} \mid l \in [m] \}}$.
We weaken this formula to obtain $\neg \beta^i_{ \{ m_{d_l,J_l} \mid l \in [m] \} } \vee \neg \beta^i_{\{ m_{{d'}_l,{J'}_l} \mid l \in [m] \}  } \vee \bigvee_{k \in I} \neg x_k$,
and from that derive  $ \neg \left(\bigwedge_{k \in I} x_k \wedge \beta^i_{ \{ m_{d_l,J_l} \mid l \in [m] \} }  \right) \vee \neg  \left(\bigwedge_{k \in I} x_k \wedge \beta^i_{ \{ m_{{d'}_l,{J'}_l} \mid l \in [m] \} }  \right)= \neg \eta_E \vee \neg \eta_F$.

\end{proof}

\section{Translations of Formulas into Polynomials}{\label{uniformity}}

\subsection{Direct Translation of Clauses}{\label{directTranslation}}

For sets of narrow clauses,  a common way to translate the clauses into
polynomials is to map $x$ to $1-x$, $\neg x$ to $x$ and replace ``OR'' by 
multiplication.  This is most commonly used for constant-width CNFs,  and in 
this case, we show that clauses efficiently reduce to their translations.

\begin{definition} \cite{Bonet:1999:SPS}
For a clause $C$ in variables ${\vec{x}}$, the {\em{direct translation of 
$C$,  $tr(C)$,}}  is defined recursively as follows:
(i) $tr(\emptyset) = 1$ (ii) $tr(A \vee x) = tr(A)(1-x)$  
(iii) $tr(A \vee \neg x) = tr(A)x$

For a CNF $F$,  the direct translation of $F$,
$tr(F)$, is the set $\{ tr(C) \mid C \in F \}$.
\end{definition}

It is easily verified by induction that an for any clause $C$, a Boolean 
assignment satisfies $C$  if and only if
it is a root of $tr(C)$.

Whenever $C$ is satisfied,  there exists an $m$-partition on the satisfied
monomials of $tr(C)$.  Moreover, if  $C$ contains at most $w$ variables,  then
the $m$-partition can be defined by  depth two formulas of size $O(2^w)$,
and by the completeness of constant-depth Frege systems,  there is a constant 
depth derivation from $C$ of size $2^{O(w)}$  that these formulas define an
$m$-partition on the satisfied monomials of $tr(C)$.  
Therefore,  $C$ reduces to $tr(C)$ in constant depth and size $O(2^w)$.

\begin{lemma}
If $F$ is an unsatisfiable CNF of $m$ clauses of width  $w$, then
$F$ is reducible to $tr(F)$ in size $m2^{O(w)}$ and depth $O(1)$.
\end{lemma}

\subsection{Translations That Use Extension Variables}

More involved translations of formulas into sets of polynomials use 
extension variables that represent sub-formulas.
The simplest way of doing this would be to reduce
an unbounded fan-in formula $\Gamma$ to a bounded fan-in formula, and
then introduce one new variable $y_g$ per gate $g$,
with the polynomial that says $y_g$ is computed correctly
from its inputs. 
It is easy to give a reduction from $\Gamma$ to this translation,
of depth $depth (\Gamma)$ and size $poly (|\Gamma|)$.
(We can define $y_g$ by the subformula rooted
at $g$ and every polynomial would have constant size,  so defining
the partition is trivial.)
However, this translation reveals little for our purposes 
because there is usually no  small degree
Nullstellensatz refutation of the resulting system
of polynomials, even for trivial $\Gamma$.  
For example, say that
we translated the formula
 $x_1,  \lnot (((((x_1 \lor x_2) \lor \cdots \lor x_n)$
this way.  The resulting system
of polynomials is weaker than the induction principles (see the 
end of this section)
which require $\Omega( \log n)$ degree {\bf NS} refutations
\cite{bussPitNullBounds}.

We give an alternative translation of formulas into sets of polynomials so
that the formula is unsatisfiable  if the set of polynomials has 
no common root.  A formula $f$ reduces to the set of polynomials with depth 
$O({\mbox{depth}}(f))$ and size $O(|f|)$.  Moreover,  for many previously
studied unsatisfiable CNFs (such as the negated counting principles),
this translation is the same as the previously studied translations
(up to constant-degree Nullstellensatz derivations).

\begin{definition}{\label{transDef}}  Let $f$ be a formula in the variables $x_1, \ldots, x_n$
and the connectives $\{ \bigvee, \neg \}$.
For each pair of subformulas $g_1$ and $g_2$ of $f$,  we write
{\em{$g_1 \rightarrow g_2$}} if $g_1$ is an input to $g_2$.
Canonically order the subformulas of $f$,  and write $g_1 < g_2$ if 
$g_1$ precedes  $g_2$ in this ordering.
For each subformula $g$ of $f$,  let there be a variable $y_g$ -
{\em{the value of $g$}}. For each pair of subformulas of $f$, 
$g_1$ and $g_2$,  so that the top connective of $g_2$ is $\bigvee$ and 
$g_1 \rightarrow g_2$, let there be a variable $z_{g_1, g_2}$-
{\em{``$g_1$ is the first satisfied input of $g_2$''}}.
The polynomial translation of $f$,  ${\mbox{POLY}}(f)$,
is the following set of polynomials:

\begin{tabbing}
123\=123\=123\=123\=123\=123\=123\=123\=123\=123\=123\=123\=\kill
 For each variable $x_i$:  \\
\> ``The  value of subformula $x_i$ is equal to $x_i$''\\
\>\> $y_{x_i}  - x_i $ \\
 For each subformula $g$ whose top connective is  $\bigvee$: \\
\>  ``if $g_1 < g_2$, $g_1 \rightarrow g$, $g_2 \rightarrow g$, and $g_1$ is satisfied , \\
\>\>  then $g_2$  is not the first satisfied input of $g$'' \\
\>\>   $y_{g_1} z_{g_2,g}$ \\
\>    ``if $g_1$ is the first satisfied input of $g$, \\
\>\>   then $g_1$ is satisfied'' \\
\>\>   $z_{g_1,g}y_{g_1} - z_{g_1,g}$ \\
\>  ``$g$ is satisfied if and only if the some input to $g$ \\
\>\>    is the first satisfied input of $g$'' \\
\>\>  $y_{g} - \sum_{g_1 \rightarrow g} z_{g_1,g} $ \\
 For each subformula $g$ whose top connective is $\neg$:  \\
\> Let $g_1$ the unique input of $g$, \\
\> ``if $g_1$ is satisfied if and only if  $g$ is not satisfied'' \\
\>\> $y_{g_1} + y_g - 1$ \\
 The formula $f$ is satisfied: \\
\>\> $y_f - 1$
\end{tabbing}

\end{definition}

One can  show  by induction  that if $f$ is satisfiable then
${\mbox{POLY}}(f)$ has a common root.
By the contrapositive,  if ${\mbox{POLY}}(f)$ has no common roots,
then $f$ is unsatisfiable. 

\begin{lemma}{\label{satToRoot}}
Let $f$ be a Boolean formula in the variables $x_1, \ldots , x_n$.
If $f$ is satisfiable,  then ${\mbox{POLY}}(f)$ has a common $0/1$ root.
\end{lemma}
\begin{proof}
Let $\alpha$ be a $0/1$ assignment to $x_1, \ldots , x_n$.
For any propositional formula $g$,  let $\alpha(g)$ denote the
value of $g$ under the assignment $\alpha$.  

Suppose that $\alpha(f)=1$.
We extend $\alpha$ to the variables of ${\mbox{POLY}}(f)$ as follows:
For each subformula $g$ of $f$,  let $\alpha(y_g) = \alpha(g)$.
When $g  = \bigvee g_i$  and $\alpha(g)=1$,
let $i_0$ be the first input to $g$ so that $\alpha(g_i)=1$.
Set $\alpha(z_{g_{i_0}})=1$ and for $i \neq i_0$, set $\alpha(z_{g_i})=0$.
When $g = \bigvee g_i$ and $\alpha(g)=0$,
$\alpha(z_{g_i,g})=0$ for all $i$.

We now show by induction that $\alpha$ is a root of ${\mbox{POLY}}(f)$.
Clearly,  for each variable $x_i$, $\alpha$ is a root of $y_{x_i} - x_i$.
Consider a subformula $\neg g$.  Because 
$\alpha(y_{\neg g})= \alpha(\neg g)$ and 
$\alpha(y_g) = \alpha(g) = 1 - \alpha(\neg g)$,  $\alpha$ is a root
of $y_{\neg g} + y_g -1$.
Consider a subformula $g = \bigvee_i g_i$.  If $\alpha(g)=0$, then
for all $i$,  $\alpha(z_{g_i,g})=0$, $\alpha(y_{g_i})=0$ and $\alpha(y_g)=0$.
In this case, $\alpha$ is clearly a root to $z_{g_i,g}y_{g_i} - z_{g_i,g}$,
$y_{g_i}z_{g_j,g}$ and $y_g - \sum_i z_{g_i,g}$.
In the case when $\alpha(g)=1$,  there exists $i_0$ so that 
$\alpha(z_{g_{i_0},g})=1$ and for all $j \neq i_0$,  $\alpha(z_{g_j,g})=0$.
Moreover, $\alpha(y_{g_{i_0}})=1$, $\alpha(y_g)=1$ and for all $j < i_0$,
$\alpha(y_{g_j})=0$. Therefore, $\alpha$ is
a root to $y_{g_j}z_{g_i,g}$ for all $i<j$, $z_{g_i,g}y_{g_i} - z_{g_i,g}$
for all $i$, and $y_g - \sum_i z_{g_i,g}$.
Finally,  $\alpha$ is a root of $y_f -1$ because $\alpha(f)=1$ by assumption.
\end{proof}

The argument of lemma \ref{satToRoot} can be carried in 
Frege systems with depth  $O(\mbox{depth}(f))$ and size $O(|f|)$.

\begin{theorem}
If $f$ is a formula in the variables 
$x_1,  \ldots, x_n$ and the connectives $\{ \bigvee, \neg \}$,
then $f$ is reducible to ${\mbox{POLY}}(f)$ in
depth $O({\mbox{depth}}(f))$ and size polynomial in $|f|$.
\end{theorem}

\begin{proof}
We proceed in two stages.  First,  we give a set of formulas, 
${\mbox{EXT}}(f)$,  that is in the variables $x_i$,  $y_g$ and $z_{g_1,g_2}$ 
and is analogous to the translation of $f$ into polynomials. 
 We show that this translation has a constant depth, polynomial size reduction
to ${\mbox{POLY}}(f)$ and then show that $f$ has a depth $O({\mbox{depth}}(f))$ reduction to
${\mbox{EXT}}(f)$ of size polynomial in $|f|$.

Let ${\mbox{EXT}}(f)$ be the following set of formulas:
\begin{tabbing}
123\=123\=123\=123\=123\=123\=123\=123\=123\=123\=123\=123\=\kill
 For each variable $x_i$:  \\
\>\> $y_{x_i}  \leftrightarrow x_i $ \\
 For each subformula $g$ whose top connective is  $\bigvee$: \\
\>  ``if $g_1 < g_2$, $g_1 \rightarrow g$, $g_2 \rightarrow g$, and $g_1$ is satisfied , \\
\>\>  then $g_2$  is not the first satisfied input of $g$'' \\
\>\>   $\neg y_{g_1} \vee \neg z_{g_2,g}$ \\
\>    ``if $g_1$ is the first satisfied input of $g$, \\
\>\>   then $g_1$ is satisfied'' \\
\>\>   $z_{g_1,g} \rightarrow  y_{g_1}$ \\
\>  ``$g$ is satisfied if and only if some input to $g$ \\
\>\>  is the first satisfied input of $g$ \\
\>\>  $y_g \leftrightarrow \bigvee_{g_1 \rightarrow g} z_{g_1,g} $ \\
 For each subformula $g$ whose top connective is $\neg$:  \\
\> Let $g_1$ the unique input of $g$, \\
\> ``if $g_1$ is satisfied then $g$ is not satisfied'' \\
\>\> $y_{g_1} \leftrightarrow \neg y_g$ \\
 The formula $f$ is satisfied: \\
\>\> $y_f$
\end{tabbing}

There is a straightforward constant-depth, polynomial-size reduction of 
${\mbox{EXT}}(f)$ to ${\mbox{POLY}}(f)$. For each polynomial
of ${\mbox{POLY}}(f)$,  there is a
formula of ${\mbox{EXT}}(f)$ that reduces to the polynomial;  the
formula associated with each polynomial is given in table V.1.
For the constant-size polynomials of ${\mbox{POLY}}(f)$,  the 
corresponding formula of ${\mbox{EXT}}(f)$ implies that there is an
$m$-partition on the satisfied variables of the polynomial.  Because the
polynomial involves a constant number of variables,  the partition may be
defined and proved correct in constant size, depth two.

\begin{table}[h]{\label{reductionTable}}
\begin{center}
\caption{Polynomials and their Associated Formulas}
\begin{tabular} {|c|c|} \hline
polynomial & associated formula  \\
\hline
$y_{x_i} - x_i$ & $y_{x_i} \leftrightarrow x_i$ \\
\hline
$y_{g_1}z_{g_2,g}$ & $\neg y_{g_1} \vee \neg z_{g_2,g}$ \\
\hline
$z_{g_1,g}  y_{g_1} - z_{g_1,g}$ & $z_{g_1,g} \rightarrow  y_{g_1}$ \\
\hline
$y_{g_1} + y_g -1$  &  $y_{g_1} \leftrightarrow \neg y_g$ \\
\hline
$y_f -1$ & $y_f$ \\
\hline
\end{tabular}
\end{center}
\end{table}

The only polynomials of ${\mbox{POLY}}(f)$ that involve a non-constant number
of variables are those of the form 
$y_{g} - \sum_{g_1 \rightarrow g} z_{g_1,g}$,  and from the hypotheses of 
${\mbox{EXT}}(f)$ it can be shown that $y_g$ is satisfied if only if
exactly one of the $z_{g_1,g}$'s is satisfied. Because there are $(m-1)$ 
copies of each $z_{g_1,g}$ in such a polynomial,  we can group $y_g$ with 
these copies of $z_{g_1,g}$ whenever $z_{g_1,g}$ is satisfied.

To reduce $f$ to ${\mbox{EXT}}(f)$,
it is easy to check to that there is a polynomial size, 
depth $O({\mbox{depth}}(f))$ derivation of the following substitution
instance of ${\mbox{EXT}}(f)$  from the hypothesis $f$.  (The substitution
instances of each formula are given in table V.2.)
\[{\mbox{EXT}}(f)[ y_g \leftarrow g, \ z_{g_1,g} \leftarrow ( g_1 \wedge \bigwedge_{g_2 < g_1 \atop g_2 \rightarrow g} \neg g_2 )] \]

\begin{table}[h]
\begin{center}
\caption{Formulas and their Substitution Instances}
\begin{tabular} {|c|c|c|} \hline
formula & substitution instance & comment\\
\hline
$y_{x_i} \leftrightarrow x_i$ &  $x_i \leftrightarrow x_i$  & \\
\hline
 $\neg y_{g_1} \vee \neg z_{g_2,g}$ &  $\neg g_1 \vee \neg ( g_2 \wedge \bigwedge_{g_3 < g_2 \atop g_3 \rightarrow g} \neg g_3 )$ &  $g_1 < g_2 $ \\
\hline
$z_{g_1,g} \rightarrow  y_{g_1}$ &  $(g_1 \wedge \bigwedge_{g_2 < g_1 \atop g_2 \rightarrow g} \neg g_2 ) \rightarrow g_1$  & \\
\hline
$y_g \leftrightarrow \vee_{g_1 \rightarrow g} z_{g_1,g} $ & 
$g \leftrightarrow \vee_{g_1 \rightarrow g} 
( g_1 \wedge \bigwedge_{g_2 < g_1 \atop g_2 \rightarrow g} \neg g_2 )$ 
& $g = \vee_{g_1 \rightarrow g} g_1$ \\
\hline
$y_{g_1} \leftrightarrow \neg y_g$ & $g_1 \leftrightarrow \neg g $ & $g = \neg g_1$ \\
\hline
 $y_f$  &  $f$ & \\
\hline
\end{tabular}
\end{center}
\end{table}
\end{proof}

{\bf Example:}  We illustrate our  translation  
with a the clauses of the negated counting principles.  The translation of
this set of clauses turns out to be same (up to constant degree Nullstellensatz
derivations) as the polynomial formulation of the 
counting principles previously studied.

Let $V$ be a set of cardinality indivisible by $m$.  The clauses are
$F_v = \bigvee_{e \ni v} x_e$ for $v \in V$ and
$G_{e,f} = \neg x_e \vee \neg x_f$ for $e,f \in [V]^m$ with $e \perp f$.
The standard translation of these systems has the polynomials
$\sum_{e \ni v} x_e$,  for $v \in V$,  and $x_e x_f$,  for $e \perp f$.

The polynomials introduced by the translation of $G_{e,f}$ are:
$y_{x_e} - x_e$, $y_{x_f}- x_f$,  $y_{\neg x_e} + y_{x_e}-1$,
$y_{\neg x_f} + y_{x_f}-1$,  $y_{\neg x_e} z_{\neg x_f, G_{e,f}}$,
$z_{\neg x_e, G_{e,f}} y_{\neg x_e} - z_{\neg x_e, G_{e,f}}$,
$z_{\neg x_f, G_{e,f}} y_{\neg x_f} - z_{\neg x_f, G_{e,f}}$,
$y_{G_{e,f}}- z_{\neg x_e, G_{e,f}} - z_{\neg x_f, G_{e,f}}$ and
$y_{G_{e,f}} -1$.   It is easy to check that thee is a constant degree
derivation of $x_e x_f$ from these polynomials (in particular,  a non-optimal
but constant-degree derivation is given by the completeness of the 
Nullstellensatz system).

The polynomials introduced by the translation of $F_v$ are:
$y_{x_e} - x_e$,  $z_{y_{x_e},F_v} y_f$ (for $e,f \ni v$ and $e<f$),
$z_{y_{x_e},F_v}y_{x_e}-z_{y_{x_e},F_v}$ (for $e \ni v$),
$y_{F_v} - \sum_{e \ni v} z_{y_{x_e},F_v}$ and $y_{F_v}-1$.
With a degree two Nullstellensatz derivation we may derive
$\sum_{e \ni v} z_{e,F_v}x_e -1$. 
Multiplying this by $\sum_{e \ni v} x_e$, and reducing using the previously
derived polynomials $x_e x_f$ and the axioms $x_e^2 - x_e$,  yields
$\sum_{e \ni v} z_{e,F_v}x_e  - \sum_{e \ni v} x_e$.  Subtracting this from
$\sum_{e \ni v} z_{e,F_v}x_e -1$ yields $\sum_{e \ni v} x_e$.

\subsubsection{A Note on Translations of Formulas to Polynomials Using
Extension Variables}

\begin{definition}
The {\em{induction principle of length $M$,
${\mbox{IND}}(M)$,}} is the 
following system of polynomials:  $y_1$, $y_{r+1}y_r - y_{r+1}$ (for $r <M$)
and $y_M - 1$. 
\end{definition}
\begin{theorem}\cite{bussPitNullBounds,homoPoly99}
The ${\mbox{IND}}(M)$ system has Nullstellensatz refutations of degree
$O(\log M)$  over any field.  Moreover,  over any field the system
requires degree $\Omega(\log M)$ Nullstellensatz refutations.
\end{theorem}

The ``standard'' translation of 
$x_n,  \lnot (((((x_n \lor x_{n-1}) \lor \cdots \lor x_1))))$ into polynomials 
using extension variables introduces new variables $z_1, \ldots, z_{n-1}$,
with   polynomials $x_n - 1$,
$1 - (1 - x_n)(1- x_{n-1}) - z_{n-1}$,
$1 - (1-z_{n-1})(1- x_{n-2}) - z_{n-2}$, \ldots,
$1 - (1 - z_2)(1-  x_1) - z_1$, and  $z_1$.   (The indices have been reversed
from those of subsection \ref{directTranslation} to ease the reduction.)

We may define this set of polynomials from ${\mbox{IND}}(n)$
using the following definitions:  $x_i := y_i$ for $i$, $1 \le i \le n$, 
and $z_i  := y_i$, for $i\le n-1$.  The polynomials $z_1 = y_1$ and 
$x_n - 1 = y_n - 1$ are belong to ${\mbox{IND}}(n)$,  and for each
$r$, $1 \le r \le n-2$, 
\[\begin{array}{lcl}
1 - (1 - z_{r+1})(1- x_r) - z_r & = & 1 - (1 - y_{r+1})(1- y_r) - y_r  \\
= 1 - (1  + y_{r+1}y_r - y_r - y_{r+1}) - y_r & = & - (y_{r+1}y_r - y_{r+1}) 
\end{array}\]

Similarly,  $1 - (1-x_n)(1-x_{n-1}) - z_{n-1} = - (y_ny_{n-1} - y_n)$.

Because there is a constant degree reduction from ${\mbox{IND}}(n)$ to
the standard translation of $x_n,  \lnot (((((x_n \lor x_{n-1}) \lor \ldots x_1))))$ into polynomials,  this translation requires super-constant
degree to refute in the Nullstellensatz system.

\section{An Application to Unsatisfiable Systems of Constant-Width Linear Equations}{\label{linEq}}

Many tautologies studied in propositional proof complexity, such as 
Tseitin's tautologies \cite{BenSasson02a}
and the $\tau$ formulas of Nisan-Wigderson generators 
built from parity functions,  can be expressed as inconsistent systems of
linear equations over a field $\zah_q$ in which each equation involves
only a small number of variables.  We show that in such situations,  
constant-depth Frege with counting axioms modulo $q$ can prove these principles
with polynomial size proofs.

Fix a prime number $q$.
Let $A$ be an $m \times n$ matrix over $\zah_q$, let $x_1, \ldots, x_n$ be 
variables and let ${\vec{b}} \in \zah_q^m$ be so that $A{\vec{x}} ={\vec{b}}$
has no solutions.  Let $w$ be the maximum number of non-zero entries in any
row of $A$.

For each $i \in [m]$,  let $A_i$ be the $i$'th row of $A$,  and let $p_i$ be
the  polynomial $A_i {\vec{x}} - b_i$.
Let $C_i$ the CNF that is satisfied if and only $p_i({\vec{x}}) = 0$.  Notice
that $C_i$ has size at most $2^w$.  The explicit encoding of 
$A{\vec{x}}={\vec{b}}$ is the CNF $\bigwedge_{i=1}^m C_i$.

The methods of subsection \ref{directTranslation} show that 
$\bigwedge_{i=1}^m C_i$ is reducible to the system of polynomials 
$\{p_1, \ldots, p_m \}$ via a constant depth reduction of size $m2^{O(w)}$.
Moreover,  the system of polynomials $\{p_1, \ldots, p_m \}$ has a degree one
Nullstellensatz refutation given by Gaussian elimination.  Moreover, degree 
one refutations are of size $O(mn)$.
Thus we have the following theorem:
\begin{theorem}{\label{lineq}}
Fix a prime number $q$.
Let $A$ be an $m \times n$ matrix, let $x_1, \ldots, x_n$ be variables
and let ${\vec{b}} \in \zah_q^m$ be so that $A{\vec{x}} ={\vec{b}}$ has
no solutions.
Let $w$ be the maximum number of non-zero entries in any row
of $A$.

There is a constant depth Frege with counting axioms modulo $q$ refutation of
the explicit encoding of $A{\vec{x}}={\vec{b}}$  of size polynomial in
$m$,$n$ and $2^w$.
\end{theorem}

The Tseitin graph tautologies on an expander graph are known to require
exponential size constant-depth Frege proofs \cite{BenSasson02a}.  Because
these principles can be represented as an unsatisfiable system of linear
equations,  they have polynomial size constant-depth Frege with counting axioms
proofs. 
\begin{corollary}
There exists a family of unsatisfiable sets of constant width clauses that
require exponential size constant-depth Frege refutations,  but have
polynomial size constant-depth Frege with counting axioms refutations.
\end{corollary}

\section{Acknowledgements}

The authors would like to thank Sam Buss for useful conversations
and insightful suggestions.

\bibliography{bib}

\end{document}